\documentclass[aps,prl,twocolumn,showpacs,superscriptaddress,showkeys]{revtex4-1}

\usepackage{amssymb}
\usepackage{amsmath}
\usepackage{amsfonts}
\usepackage{graphicx}
\usepackage{color}
\usepackage{xspace}
\usepackage{ulem}
\usepackage{mathtools}
\usepackage{hhline}
\usepackage{tikz}

\usepackage{dcolumn}
\usepackage{bm}
\usepackage{hyperref}
\usepackage{comment}

\usepackage[hyperref,svgnames]{xcolor}

\newcommand{\gsim}{\lower.7ex\hbox{$\;\stackrel{\textstyle>}{\sim}\;$}}
\newcommand{\lsim}{\lower.7ex\hbox{$\;\stackrel{\textstyle<}{\sim}\;$}}

\def\beq{\begin{equation}}
\def\eeq{\end{equation}}
\def\bea{\begin{eqnarray}}
\def\eea{\end{eqnarray}}
\def\bitem{\begin{itemize}}
\def\eitem{\end{itemize}}
\newcommand{\bec}{\begin{center}}
\newcommand{\eec}{\end{center}}
\newcommand{\ba}{\begin{array}}
\newcommand{\ea}{\end{array}}

\def\bar#1{\overline{#1}}

\def\inv{^{\raise.15ex\hbox{${\scriptscriptstyle -}$}\kern-.05em 1}}
\def\lbar{{\lower.35ex\hbox{$\mathchar'26$}\mkern-10mu\lambda}} %lambda bar

\let\<=\langle
\let\>=\rangle

\let\+=\uparrow

\def\beqa{\begin{equation}\begin{aligned}}
\def\eeqa{\end{aligned}\end{equation}}

\newcommand{\solidblue}{%
\begin{tikzpicture}[baseline=-0.6ex]
\draw[blue, line width=1.5pt, line join=round] (0,0) -- (1.1,0);
\end{tikzpicture}}
\newcommand{\dashedblue}{%
\begin{tikzpicture}[baseline=-0.6ex]
\draw[blue, line width=1.5pt, line join=round, dash pattern=on 3.0pt off 1.0pt] (0,0) -- (1.1,0);
\end{tikzpicture}}
\newcommand{\solidred}{%
\begin{tikzpicture}[baseline=-0.6ex]
\draw[red, line width=1.5pt, line join=round] (0,0) -- (1.1,0);
\end{tikzpicture}}
\newcommand{\dashedred}{%
\begin{tikzpicture}[baseline=-0.6ex]
\draw[red, line width=1.5pt, line join=round, dash pattern=on 3.0pt off 1.0pt] (0,0) -- (1.1,0);
\end{tikzpicture}}

\newcommand{\AddrSlovenia}{Jožef Stefan Institute, Jamova 39, 1000 Ljubljana,  Slovenia}
\newcommand{\AddrCoimbra}{Univ Coimbra, Faculdade de Ci\^encias e Tecnologia da Universidade de Coimbra and CFisUC, Rua Larga, 3004-516 Coimbra, Portugal}

\begin{document}

\title{Lukewarm inflation, primordial black holes and gravitational waves}

\author{Paulo B. Ferraz}\email{paulo.ferraz@student.uc.pt}\affiliation{\AddrCoimbra}
\author{Ant\'onio Torres Manso}\email{antonio.torres.manso@ijs.si}\affiliation{\AddrCoimbra}\affiliation{\AddrSlovenia}
\author{Jo\~{a}o G.~Rosa} \email{jgrosa@uc.pt}\affiliation{\AddrCoimbra}

\date{\today}

\begin{abstract}
We show that a secondary period of warm inflation, at temperatures parametrically below the GUT scale, may not only dilute any unwanted thermal relics formed after reheating but also significantly enhance the power spectrum of primordial curvature perturbations on small scales. The latter is nearly scale-invariant over an exponentially large range of comoving scales, resulting in a stochastic gravitational wave background with a nearly constant energy density over a broad range of frequencies and an associated population of sub-solar mass primordial black holes (PBHs). This scenario could, in particular, explain the NANOGrav signal if it persists at much higher frequencies with a comparable magnitude, up to a sharp cut-off, $f_{max}$. The associated PBH mass distribution may exhibit either a peak at a mass $M_*\propto f_{max}^{-2}$ or a broad plateaux above this mass threshold, depending on the spectral tilt. Moreover, in this scenario the NANOGrav signal is compatible with a significant fraction of dark matter in PBHs in the asteroid- and planetary-mass ranges.
\end{abstract}

%\pacs{} 

\maketitle

%%%%%%%%%%%%%%%%%%%%%%%%%%%%%%%%%%%%%%%%%%%%%%%%%%%%%%%%%%%%%%%%%%%%%%%%%%%%%%%%%%%%%%%%

%%%%%%%%%%%%%%%%%%%%%%%%%%%%%%%%%%%%%%%%%%%%%%%%%%%%%%%%%%%%%%%%%%%%%%%
%%%%%%%%%%%%%%%%%%%%%%%%%%%%%%%%%%%%%%%%%%%%%%%%%%%%%%%%%%%%%%%%%%%%%%%
%%%%%%%%%%%%%%%%%%%%%%%%%%%%%%%%%%%%%%%%%%%%%%%%%%%%%%%%%%%%%%%%%%%%%%%
%%%%%%%%%%%%%%%%%%%%%%%%%%%%%%%%%%%%%%%%%%%%%%%%%%%%%%%%%%%%%%%%%%%%%%%

\section{Introduction}

The inflationary paradigm \cite{inflation} is currently the most widely favoured explanation for the nearly-scale invariant primordial spectrum of super-horizon curvature perturbations inferred from Cosmic Microwave Background (CMB) and Large Scale Structure (LSS) observations. Its simplest implementation considers a single scalar field following a ``slow-roll'' trajectory down a very flat potential, sustaining roughly 50-60 e-folds of accelerated expansion in the early universe. While this hypothesis seems compatible with observations so far, it is not entirely free of theoretical problems, from the required fine-tuning of the scalar potential (the so-called ``eta-problem'') to the large temperatures generically attained in the subsequent reheating process as the inflaton's energy was converted into radiation. Such large temperatures may have resulted in the thermal production of unwanted relics predicted in several beyond the Standard Model (BSM) scenarios, like gravitinos, scalar moduli or topological defects associated with post-inflationary phase transitions.

Since CMB and LSS observations can only probe perturbations in a relatively small range of comoving scales ($\lesssim 10$ e-folds), we may alternatively consider scenarios with two (or more) periods of inflation, driven by different scalar fields. There is, in fact, no shortage of inflaton candidates in the most promising SM extensions (supersymmetry, extra-dimensions, string theory, multi-Higgs, etc). The observed CMB temperature fluctuations would then have been seeded during a primary inflation period, with the secondary period diluting away any thermal relics produced during the first reheating process.  One of the first proposals along these lines was {\it thermal inflation}, based on a scalar flat direction temporarily stuck in a metastable minimum by thermal effects. Despite its theoretical appeal, thermal inflation is hard to probe with astrophysical observations, since it generates relatively small curvature perturbations on small scales \cite{Dimopoulos:2019wew,Bastero-Gil:2023sub,Bae:2025vkn}.

In this Letter, we propose an alternative approach to a secondary inflationary period based on the dynamics of warm inflation, i.e.~a slow-roll trajectory for a scalar field that is sustained by thermal friction, due to its interactions with a nearly-thermal radiation bath \cite{Berera:1995ie,Berera:1995wh}. Warm inflation scenarios have been widely studied in the literature, but always aiming to explain the spectrum of primordial curvature perturbations on CMB scales, which generically requires temperatures close to the grand unification theory (GUT) scale $\sim 10^{14}-10^{15}$ GeV. Here, we will show that thermal friction effects are naturally much stronger during a secondary period of {\it lukewarm inflation}, at temperatures parametrically below the GUT scale. This leads to a significant enhancement of the curvature perturbation power spectrum on small scales, compared to those that leave the horizon during the primary inflation stage. Lukewarm inflation may consequently leave distinctive observational signatures that we discuss in detail below, namely a stochastic gravitational wave background (SGWB) and the associated formation of sub-solar mass primordial black holes (PBHs).

The dynamics of lukewarm inflation is, analogously to standard warm inflation scenarios, described by the effective Langevin equation followed by the scalar inflaton field, taking into account the leading thermal effects of the nearly-thermal radiation bath it interacts with:
\begin{equation} \label{Langevin}
\ddot\phi + 3H\dot\phi +\Upsilon\dot\phi+ V'(\phi)=\xi~,
\end{equation}
 where $H$ is the Hubble parameter, $\Upsilon$ and $V(\phi)$ are the finite-temperature dissipation coefficient and scalar potential, respectively, and $\xi$ the Gaussian white noise-term resulting from the random interactions between the scalar field and the radiation bath. Note that non-local dissipation effects are well approximated by the local friction term $\Upsilon\dot\phi$ when the scalar field evolves on time scales much longer than the radiation bath's relaxation time, which is the case in the slow-roll regime. If the system remains close to thermal equilibrium, with a slowly evolving temperature $T\gg H$, the fluctuation-dissipation theorem yields ${\langle\xi_k(t)\xi_{k'}(t')\rangle=2\Upsilon T a^{-3} (2\pi)^3\delta^3(k-k')\delta(t-t')}$ in the strong dissipation regime $\Upsilon \gg H$ that we are interested in.
 
 Dissipation results in energy transfer between the inflaton field and the radiation bath through particle production, with the radiation energy density then satisfying:
\begin{equation} \label{radiation}
\dot\rho_R+4H\rho_R=\Upsilon\dot\phi^2~,
\end{equation}
 as can easily be obtained from the covariant conservation of the total energy-momentum tensor. In the slow-roll regime, Eqs.~(\ref{Langevin}) and (\ref{radiation}) for the background field and radiation energy density reduce to:
\begin{equation} \label{slow_roll}
3H(1+Q)\dot\phi \simeq -V'(\phi)~, \qquad 4H\rho_R=\Upsilon\dot\phi^2~,
\end{equation}
 where $Q=\Upsilon/3H \gg 1$. Such a trajectory leads to accelerated expansion provided that ${\epsilon_H = -\dot{H}/H^2\simeq \epsilon_\phi/(1+Q)< 1}$, alongside ${\eta_\phi/(1+Q)< 1}$. Strong dissipation effects may thus sustain a period of slow-roll inflation at finite temperature even for a heavy inflaton scalar field, $m_\phi\gg H$. These equations can be combined to yield $\rho_R/\rho_\phi\simeq \epsilon_H/2$ for $Q\gg 1$, thus showing that radiation is sub-dominant but not necessarily negligible during the slow-roll period, generically becoming the dominant component at the end of lukewarm inflation when $\epsilon_H\gtrsim 1$ with no need for a subsequent reheating period.
 
  This, of course, neglects thermal corrections to the inflaton mass, which is only possible for particular forms of the interactions between the inflaton and the fields in the radiation bath. Here we will consider the {\it Warm Little Inflaton} (WLI) setup \cite{Bastero-Gil:2016qru,Levy:2020zfo, Ferraz:2023qia}, in which the inflaton arises from the collective breaking of a U(1) gauge symmetry, corresponding to the relative phase between two complex Higgs fields. In its original version, the inflaton interacts with two fermion fields, which are naturally light ($m\ll T$) in this construction, and their leading thermal corrections to the scalar potential are cancelled via a discrete interchange symmetry. It was later shown that thermal corrections are under control due to their oscillatory nature even without imposing the latter symmetry \cite{Ferraz:2023qia}. In this setup the dissipation coefficient is simply proportional to the radiation temperature, $\Upsilon = C_T T$, where the proportionality constant $C_T\sim g^2/h^2\lesssim 1$ depends on two Yukawa couplings $g$ and $h$ that control the inflaton-fermion interactions and the fermion thermal decay width, respectively \cite{footnote1}.
 In this case we have:
\begin{equation} \label{Q}
Q = {C_T\over 3}\left({T\over H}\right) \simeq \left({5\over \pi^2}{\epsilon_H\over g_*}\right)^{1/2} C_T{M_P\over T} ~,
\end{equation}
 where $M_P\simeq 2.4\times10^{18}$ GeV is the reduced Planck mass and $g_*$ denotes the number of relativistic species in the radiation bath. Hence, if lukewarm inflation occurs at low temperatures, particularly below the GUT scale, dissipation effects are naturally very large, $Q\gg 1$, assuming $C_T\lesssim 1$, i.e.~that there is no unnaturally large hierarchy between the dimensionless couplings $g$ and $h$.
 
 This not only makes lukewarm inflation easier to realize than conventional warm inflation models at the GUT scale, with only weak or mild dissipative effects sustaining the slow-roll regime, but also has a profound impact on the evolution of curvature perturbations. In warm inflation, inflaton perturbations are sourced by its interactions with the radiation bath both directly, through the noise term in Eq.~(\ref{Langevin}), and indirectly, through the temperature dependence of the dissipation coefficient. While there are scenarios where $\Upsilon$ decreases with temperature in certain parametric regimes \cite{Bastero-Gil:2019gao}, in most models (including the WLI) dissipative effects are stronger at higher temperatures, noting also that generically $\Upsilon \rightarrow 0$ as $T\rightarrow 0$ in the adiabatic regime \cite{Moss:2008lkw}. This means that in regions with a slightly larger temperature thermal friction transfers more energy into the radiation bath, heating it up more than initially colder regions. This amplifies temperature fluctuations and the inflaton perturbations to which they are coupled to, leading to a {\it growing mode} in the primordial curvature power spectrum. Overall, the (luke)warm primordial curvature power spectrum on super-horizon scales is given by, for $Q\gg1$ \cite{Graham:2009bf,Bastero-Gil:2011rva,Ramos:2013nsa}:
\begin{equation} \label{power_spectrum}
\Delta_\mathcal{R}^2= {\sqrt{3\pi}\over 24\pi^2}{V\over M_P^4}{Q^{3/2}\over \epsilon_H}{T\over H} G(Q)~,
\end{equation}
 where all quantities are evaluated when the relevant comoving scale crosses the horizon, $k=aH$. The function $G(Q)$ encodes the effects of the growing mode and has been determined by numerically solving the coupled equations for the inflaton and radiation fluctuations, growing as a power law, $G(Q)\propto Q^\alpha$, for $Q\gg 1$.  Using the numerical package WI2easy \cite{Rodrigues:2025neh}, we have obtained $G(Q)\simeq 5\times10^{-4}\, Q^3$ and agreement with other parameterizations of this function valid for only mildly strong dissipative effects. Since all relevant quantities follow a slow-roll evolution, this yields a nearly scale-invariant primordial power spectrum $\Delta_{\mathcal{R}}^2= A_s (k/k_{max})^{n_s-1}$ for $k_{min}<k<k_{max}$, where $k_{min/max}$  denotes the first/last scale to become super-horizon during lukewarm inflation, with an amplitude (evaluated at $k_{max})$:
\begin{equation} \label{power_spectrum_A}
A_s\simeq 0.07 \left({100\over g_*}\right)^{7/4}\left({C_T\over 10^{-2}}\right)^{9/2}\left({ 10^7\ \mathrm{GeV}\over T_R}\right)^{3/2}~,
\end{equation}
where $T_R$ denotes the ``reheating'' temperature at the end of the slow-roll regime (and onset of radiation domination), $\epsilon_H=1$, in general comparable to the initial temperature of lukewarm inflation.
  
 We thus see that the amplitude of primordial curvature perturbations decreases with the temperature at which lukewarm inflation occurs.  Interestingly, note that the inferred value on CMB scales, $A_s\sim 10^{-9}$ corresponds to temperatures parametrically close to the GUT or Planck scale, $T_R\sim 17M_P(100/g_*)^{7/6}C_T^3$, depending on the strength of the dissipation coefficient, thus suggesting that a high-temperature primary stage of warm inflation model could naturally explain the amplitude of CMB fluctuations. Irrespectively of this, a secondary period of lukewarm inflation at low-temperatures naturally yields large curvature perturbations on small scales (unless of course $C_T$ is very small, which requires a large hierarchy between the couplings involved as explained above).
 
The spectral index $n_s$ depends on the scalar potential, which in the WLI setup can take an arbitrary form \cite{footnote2}. Using the slow-roll Eqs.~(\ref{slow_roll}) one obtains, for $Q\gg1$:
\begin{equation} \label{power_spectrum_A}
n_s-1\simeq {d\log \Delta_\mathcal{R}^2\over dN_e}\simeq {3\over 2}{Q'\over Q}= {3\over 10}{6\epsilon_\phi-2\eta_\phi\over Q}~,
\end{equation}
where primes denote derivatives with respect to the number of e-folds. This shows that lukewarm inflation generically predicts a nearly scale-invariant curvature perturbation power spectrum. Curiously, the spectrum is blue-tilted (red-tilted) for the same forms of the scalar potential that yield a red-tilted (blue-tilted) in single-field cold inflation scenarios. On the one hand, for the simplest forms like monomial potentials, deviations from scale invariance decrease with the duration of the lukewarm inflation stage. For instance, for a quadratic potential, $V(\phi)\propto \phi^2$, we obtain $n_s-1\simeq (N_e+5/6)^{-1}$. On the other hand, there are several scenarios, e.g.~hill-top or hybrid potentials $V(\phi)\simeq V_0\pm {1\over 2}m^2\phi^2$, where $n_s$ and $N_e$ are not directly linked. Note that the WLI setup accommodates different forms of $V(\phi)$, given that the inflaton field, being a relative phase, is U(1) gauge-invariant. In our discussion henceforth we will take $n_s$ and $N_e$ as independent parameters, considering examples with either $n_s>1$ or $n_s<1$ to illustrate the possible observational signatures of lukewarm inflation.

There are then two main observational signatures of lukewarm inflation. First, large curvature perturbations source tensor perturbations at second order in perturbation theory. This leads to a SGWB with a present abundance given by:
%
%\begin{eqnarray} \label{SGWB}
%\Omega_{GW}(k)h^2\simeq {3\times10^{-6}\over g_{*c}^{1/3}}\!\!\!
%\int_0^\infty \!\!\!dv \!\int_{|1-v|}^{1+v}\!\!\!\!\!\!du\, S(u,v,k\tau_c)
%\left({4v^2-(1+v^2-u^2)^2\over 4uv}\right)^2\Delta_\mathcal{R}^2(kv)\Delta_\mathcal{R}^2(ku)\bar{I_{RD}^2}(u,v,k\tau_c)
%\end{eqnarray}
% 
\begin{eqnarray} \label{SGWB}
\Omega_{GW}(k)h^2\simeq {3\times10^{-6}\over g_{*c}^{1/3}}\!\!\!
\int_0^\infty \!\!\!dv \!\int_{|1-v|}^{1+v}\!\!\!\!\!\!du\, \bar{S}_c\Delta_\mathcal{R}^2(kv)\Delta_\mathcal{R}^2(ku)
%\left({4v^2-(1+v^2-u^2)^2\over 4uv}\right)^2\Delta_\mathcal{R}^2(kv)\Delta_\mathcal{R}^2(ku)\bar{I_{RD}^2}(u,v,k\tau_c)
\end{eqnarray}
%
%\begin{eqnarray} \label{SGWB2}
%S(u,v,k\tau_c)&=&\left({4v^2-(1+v^2-u^2)^2\over 4uv}\right)^2\Delta_\mathcal{R}^2(kv)\Delta_\mathcal{R}^2(ku)\nonumber\\
%&\times &\bar{I_{RD}^2}(u,v,k\tau_c)
%\end{eqnarray}
%
where an analytical approximation for the time averaged function $\bar{S}_c \equiv \bar{S}(u,v,k\tau_c)$ is given in \cite{Kohri:2018awv}, 
with quantities evaluated at conformal time $\tau_c$ when the perturbations have reentered the horizon during the radiation era that follows lukewarm inflation. The resulting GW spectrum is thus nearly scale-invariant (either red- or blue-tilted depending on $n_s$) in the frequency range corresponding to $k_{min}<k<k_{max}$, with $\Omega_{GW}\propto f^{2(n_s-1)}$. In particular, it exhibits a sharp cut-off at 
$f_{max} \simeq 2.6\times 10^{-8}(T_R/\mathrm{GeV})(g_*/100)^{1/6}$ Hz, while below  $f_{min}\simeq e^{-N_e}f_{max}$ it follows the $f^3$ infrared behavior common to a wide range of sources.
 
\begin{figure}[h]
 \includegraphics[scale=0.41]{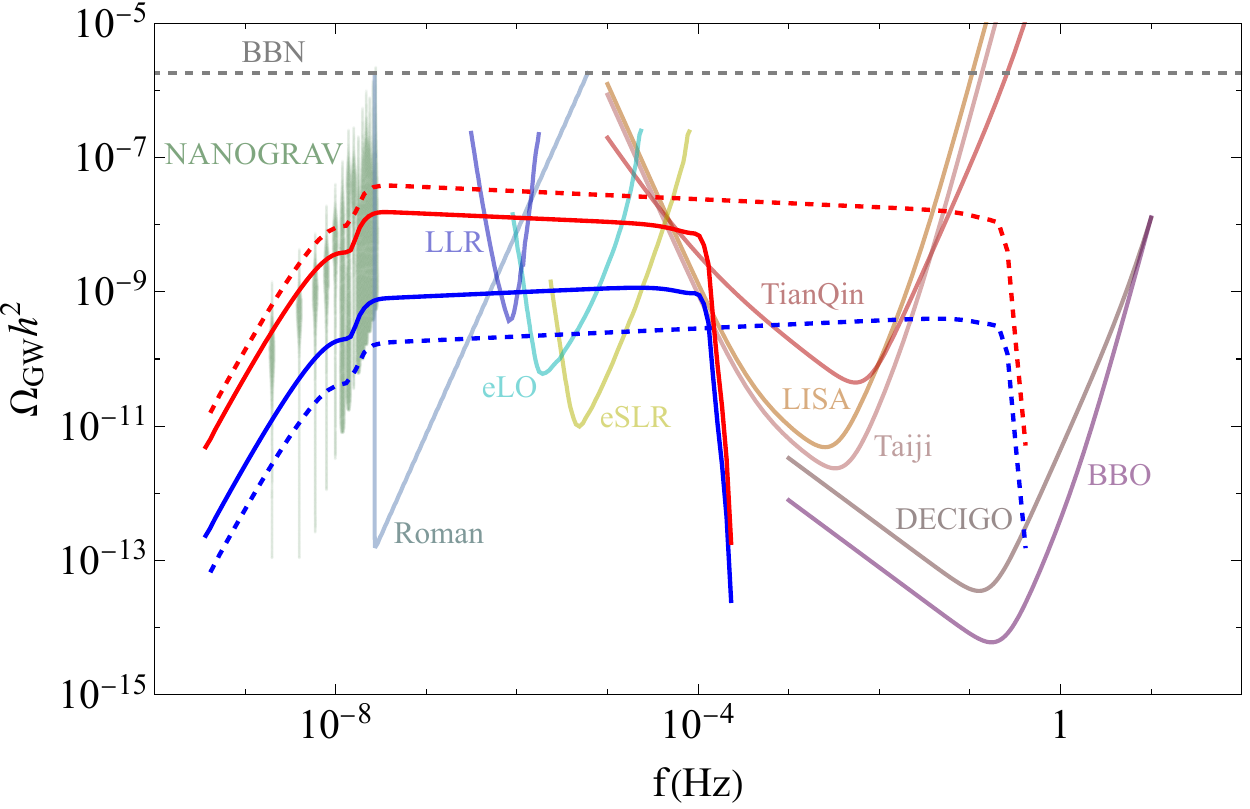}
     \begin{tabular}{c c c c c} 
        \hline 
        $f(M)\quad$ & $T_R/\mathrm{TeV}\quad$ 
        & $N_e\quad$ 
        & $n_s\quad$ 
        & $C_T$ \tabularnewline
        \hline
        \solidblue $\quad$& $5 \quad$& $9\quad$ & $1.03\quad$ & $6.9\times10^{-4}$ \\
        \solidred $\quad$& $5\quad$& $9\quad$ & $0.97\quad$ & $6.7\times10^{-4}  $  \\
        \dashedblue $\quad$& $10^4\quad$ & $17\quad$ & $1.03\quad$ & $8.5\times10^{-3}$  \\
        \dashedred $\quad$& $10^4\quad$ & $17\quad$ & $0.97\quad$ & $8.1\times10^{-3} $  \\
        \hline 
    \end{tabular}
\caption{SGWB spectrum for different realizations of lukewarm inflation. 
The  green violins represent the NANOGRAV signal \cite{NANOGrav:2023ctt}.  Future observational curves including the  Roman Telescope \cite{Wang:2022sxn,Pardo:2023cag}, the 
binary resonances \cite{Foster:2025nzf,Blas:2026xws} with 
LLR, eLO,  and eSLR, LISA \cite{LISA:2024hlh}, Taiji \cite{Luo:2019zal},
and TianQin \cite{Luo:2025ewp} are also represented in the figure. The gray-dashed line is the bound from relativistic degrees of freedom  at BBN. We have taken $g_*=100$ in our examples.
\label{gws} }
 \includegraphics[scale=0.32]{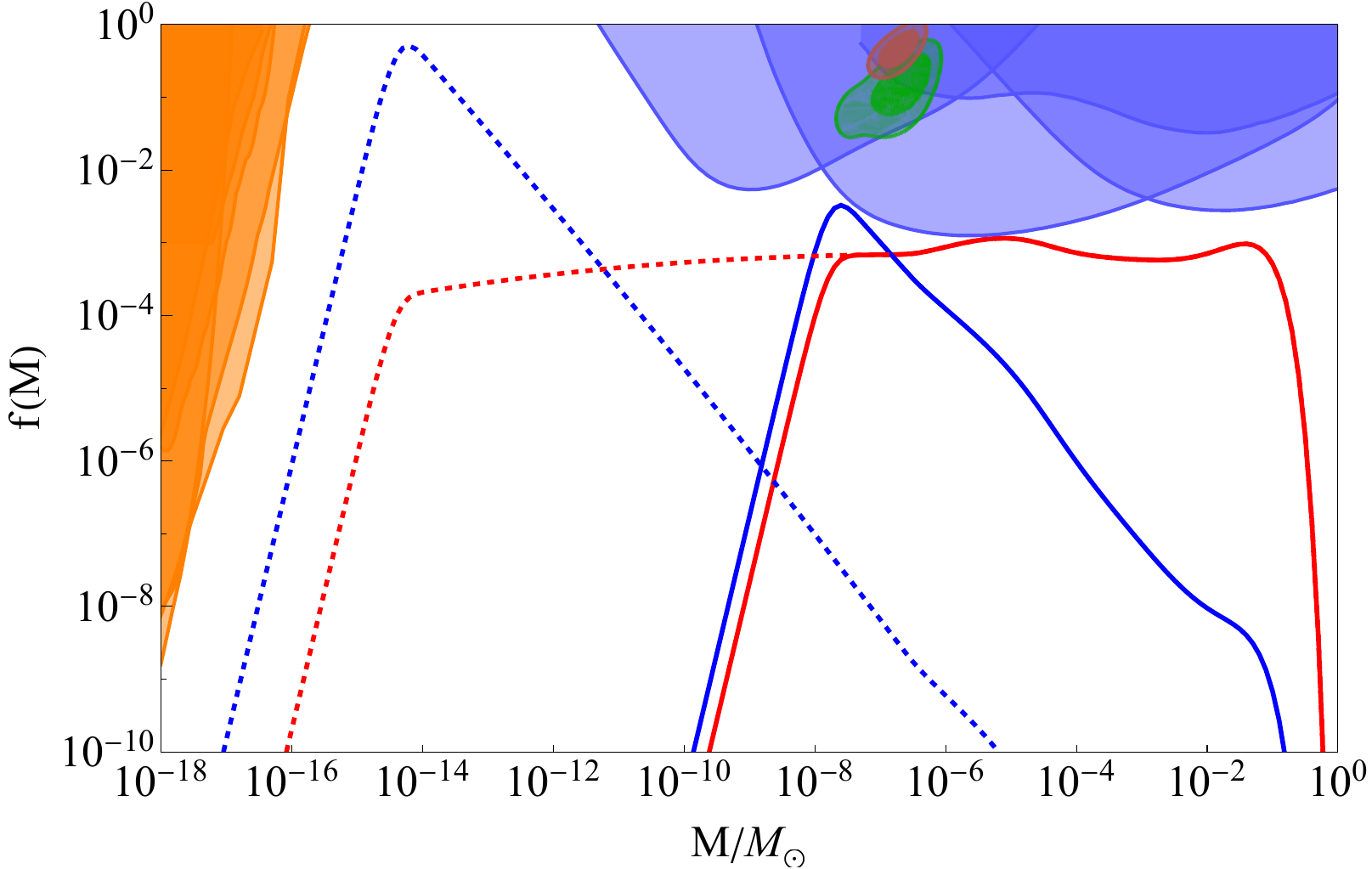}
 \caption{Fraction of dark matter in PBHs as a function of their mass $M$, for the same lukewarm inflation scenarios considered in Fig.~\ref{gws}.   The red and green shaded contours represent the allowed regions of PBH
abundance when all 12 microlensing candidates or the 4 secure candidates (“S4”) of Subaru-HSC are attributed to PBHs, respectively \cite{Sugiyama:2026kpv}.
 The shaded regions give the current constraints on the PBH abundance from Hawking evaporation \cite{Carr:2009jm, Acharya:2020jbv, Chluba:2020oip, Carr:2016hva, Boudaud:2018hqb} (orange) and gravitational microlensing \cite{Smyth:2019whb, Niikura:2019kqi, Zumalacarregui:2017qqd, Mroz:2024wia} (blue). Data was obtained from \cite{BJKavanagh}.}  

\label{pbhs}
 \end{figure}

 In Fig.~\ref{gws} we illustrate the form of the SGWB spectrum for two different values of the temperature $T_R$ and of the spectral index $n_s$. We have chosen the number of e-folds of the lukewarm inflation stage and the amplitude of the dissipation coefficient $C_T$ in each case to approximately match the signal reported by Pulsar Timing Array (PTA) collaborations in the $1-10$ nHz range (also taking into account the PBH abundance discussed below).

 The second observational implication of the naturally large curvature perturbations in lukewarm inflation is PBH formation, following the gravitational collapse of overdense regions in the subsequent radiation-dominated era, upon horizon reentry. While more sophisticated methods should be used to precisely compute the resulting PBH abundance, namely adapting peak theory methods to the particular features of lukewarm inflation, here we will employ the Press-Schechter formalism to illustrate the main features of the PBH mass function and its dependence on the model's parameters. In this formalism, a region with density contrast $\delta\equiv \delta\rho/\rho \simeq -(4/9)(aH)^{-2}\nabla^2\mathcal{R}>\delta_c$ collapses into a PBH of mass $M\simeq \mathcal{K} M_H(\delta-\delta_c)^\gamma$, where $M_H$ is the mass within the Hubble radius when the perturbation reenters the horizon, $k=aH$, with a critical exponent $\gamma\simeq 0.36$ and $\mathcal{K}\simeq 4$. The critical density contrast $\delta_c$ depends on both the shape of the perturbation \cite{Germani:2018jgr,Musco:2020jjb} and the equation-of-state parameter $w$ \cite{Harada:2013epa,Escriva:2020tak}. Here we take the value $\delta_c=0.49$ for $w=1/3$ in a pure radiation-era but consider the more realistic parameter $w(T)$ determined in \cite{Byrnes:2018clq,Carr:2019kxo}, and which takes into account its reduction during the electroweak and QCD crossovers, as well as $e^+e^-$ annihilation.
 
 For a Gaussian curvature power spectrum, the fraction of dark matter in PBHs of mass $M$ is then given by:
\begin{eqnarray} \label{PBH_spectrum}
f_{PBH}(M)={2\mathcal{K}/\gamma\over \Omega_{DM}}\!\int {dM_H \over M_H} \sqrt{M_{eq}\over M_H}{\mu^{{1\over\gamma}+1}\over \sqrt{2\pi\sigma^2}}e^{-{\left(\mu^{1\over\gamma}+\delta_c\right)^2\over 2\sigma^2}},\nonumber\\
\end{eqnarray}
%\begin{eqnarray} \label{PBH_spectrum}
%f(M)={2\over \Omega_{DM}}\!\int d\ln M_H \sqrt{M_{eq}\over M}{M\over \gamma M_H}{\mu^{1\over\gamma}\over \sqrt{2\pi\sigma^2}}e^{-{\left(\mu^{1\over\gamma}+\delta_c\right)^2\over 2\sigma^2}}\nonumber\\
%\end{eqnarray}
%
where $\mu\equiv M/(\mathcal{K}M_H)$, $M_{eq}\simeq 2.8\times 10^{17}M_\odot$ is the horizon mass at matter-radiation equality and $\sigma^2(M_H)=\int d\ln k \Delta^2_\delta (k) e^{-{k^2\over a^2H^2}}$ is the variance of the density perturbations smoothed over the Hubble radius using a Gaussian window function, with $(aH)^{-1}= k_{eq}^{-1}\sqrt{M_H/M_{eq}}$ \cite{footnote3}. 
In Fig.~\ref{pbhs}, we illustrate the PBH mass function corresponding to the scenarios considered in Fig.~\ref{gws}.  In all cases the total contribution of PBHs to the dark matter density, $f_{PBH}^{total}=\int d\ln M f_{PBH}(M)\leq 1$,  with $f_{PBH}^{total}\simeq1$ for the dashed blue curve.

On the one hand, at small masses, the PBH mass function  $f_{PBH}(M)\propto M^{1+1/\gamma}$, a universal behavior characteristic of critical collapse and insensitive to the form of the curvature power spectrum.
On the other hand, for large PBH masses we find approximately that $f_{PBH}(M)\propto M^{-{1\over 2}-{a\over 2}(n_s-1)}$, where $a\equiv {81\delta_c^2\over 16A_s}$, as shown in detail in the appendix for the interested reader. While this is in agreement with the $M^{-1/2}$ scaling obtained in \cite{Domenech:2026nun} for an exactly scale-invariant curvature power spectrum ($n_s=1$), this shows that small deviations from scale invariance may have a significant impact on the shape of the PBH mass function, taking into account that  $a\gtrsim 1$ for $A_s\sim 10^{-2}$. In the examples shown in Fig.~\ref{pbhs}, $a\sim 30$, and we find completely different PBH mass distributions for red- and blue-tilted spectra. While for $n_s>1$ we find that $f_{PBH}(M)$ is sharply peaked, scenarios with $n_s<1$ may exhibit a plateaux in the PBH mass function over several mass decades.

The transition between the two distinct behaviors of the PBH mass function occurs at ${M_{\star}\simeq 5\times 10^{-7}M_\odot\left(100/g_*\right)^{1/2}\left(T_R/\textrm{TeV}\right)^{-2}}$. This coincides with a peak in the mass function for $n_s\gtrsim 1-a^{-1}$, and the $n_s$-dependent power law behavior extends until $M_{max}\sim e^{2N_e}M_*$. We note that non-Gaussian effects may have an impact on the PBH mass function, since $f_{NL}\simeq 8$ for $\Upsilon\propto T$ in the strong dissipative regime relevant to lukewarm inflation \cite{Moss:2011qc,Bastero-Gil:2014raa}, and we plan to refine our analysis to take this into account in future work.

 Our results nevertheless show that lukewarm inflation has the potential to generate all, or at least a significant fraction of, the dark matter in sub-solar mass PBHs, namely in the asteroid ($10^{-16}-10^{-12}M_\odot$) and planetary mass windows. The latter is particularly relevant given current potential hints for gravitational microlenses with $10^{-7}-10^{-6}M_\odot$ found by HSC-Subaru from observations of M31 \cite{Sugiyama:2026kpv}, which could potentially be explained by PBHs in this mass range. Although this seems to be in tension with OGLE observations of the Magellanic Clouds \cite{Mroz:2024wia, Mroz:2026nez} shown in Fig.~\ref{pbhs}, we note that a slightly larger value ($\simeq 3\%$)  of the dissipation constant $C_T$ than the one chosen in this figure for $T_R=5$ TeV and $n_s>1$ could easily explain the HSC-Subaru microlensing events, also potentially agreeing with the PTA signal. It is interesting to note a lukewarm inflation scenario at these temperatures would signal new physics just above the energy scale currently being probed at the LHC.

Lukewarm inflation thus leads to two independent observational signatures, both with distinctive features that reflect the underlying large-amplitude and nearly-scale invariant spectrum over an exponentially large range of scales. The amplitude and unique shapes of $\Omega_{GW}(f)$ and $f_{PBH}(M)$ provide, as described above, two independent ways of determining the four parameters that characterize the primordial curvature power spectrum ($A_s, n_s, k_{min}$ and $k_{max}$), and the underlying properties of the lukewarm inflation stage ($T_R, N_e, C_T$ and the potential slow-roll parameters). We may, in particular, establish consistency relations between observables that allow us to determine whether such a period occurred in the early universe or not. For instance, we find:
\begin{eqnarray}
    M_\star \simeq 10^{-10}M_\odot\left({g_*\over 100}\right)^{-1/6}\left(\frac{\mathrm{mHz}}{f_{max}}\right)^2~,
\end{eqnarray}
with only a very mild dependence on the number of relativistic species. As discussed above, the slopes of $\Omega_{GW}(f)$ and $f_{PBH}(M)$ at large frequencies/PBH masses are also related to each other and to the amplitude of the curvature power spectrum $A_s$. In addition, we have $f_{max}/f_{min}\simeq (M_{max}/M_*)^{1/2}\simeq e^{N_e}$ (up to small model-dependent changes in the Hubble parameter during lukewarm inflation).

As shown in Fig.~\ref{gws}, the NANOGrav signal could potentially be explained by a period of lukewarm inflation, although this is also true for several other GW sources exhibiting the same $\sim f^3$ infrared behavior. A `smoking-gun' for lukewarm inflation would be a signal that persists with a nearly constant amplitude in the frequency ranges probed by planned GW detectors such as Lunar Laser Ranging (LLR) or LISA {\it and} exhibits a sharp high-frequency cut-off ($f_{max}$) \cite{footnote4}. Moreover, from such a signal we could deduce the abundance and mass distribution of the associated PBH population.

In a companion paper we describe a concrete model of lukewarm inflation (in a WLI-like setup) within a simple extension of the Standard Model with three right-handed neutrinos and the scalar fields generating their Majorana masses (one of them acting as the inflaton). We show, in particular, that a lukewarm inflation period around the QCD scale could dilute the abundance of the QCD axion, thus alleviating the fine-tuning of scenarios with axion decay constants close to the GUT scale, at the same time decreasing an initially large baryon-to-entropy ratio to the value inferred from CMB observations and the abundance of light nuclei. This model has the additional feature of predicting an early-matter era after lukewarm inflation dominated by the lightest right-handed neutrino, before it decays into Standard Model degrees of freedom. This should have an impact on PBH formation that we plan to investigate in future work, since the resulting decrease in pressure lowers the threshold $\delta_c$ for PBH formation, as well as potentially leading to PBHs with non-negligible spins \cite{Harada:2017fjm}. We note that the same particle physics setup can  realize lukewarm inflation at temperatures from a few hundred MeV up to the GUT scale, the post-inflationary dynamics depending on the model parameters.

In summary, lukewarm inflation dilutes the abundance of unwanted thermal relics produced after the first stage of inflation, can be realized within consistent particle physics setups and has unique observational signatures that can potentially be probed with future experiments.

%Lukewarm inflation is thus an appealing scenario both from the theoretical and observational perspectives, providing a secondary stage of inflation that could dilute unobserved thermal relics generated after the first inflation period and modify the spectrum of curvature perturbations at small scales in a testable way. This spectrum is characterized by four quantities ($A_s$, $n_s$, $k_{min}$ and $k_{max}$) linked to the temperature and duration of the lukewarm inflation period ($T_R$ and $N_e$), the strength of thermal friction effects parametrized by $C_T$ and the shape of the scalar potential determining the spectral tilt. It generically predicts a nearly (but not exactly) scale-invariant SGWB spectrum over an exponentially large range of frequencies, which can be further probed by future detectors such as those shown in Fig.~\ref{gws}, features that also translate into the associated mass distribution of sub-solar PBHs. Our results clearly show that lukewarm inflation may provide a simple and natural explanation for the reported PTA and gravitational microlensing signals, and we plan to perform a detailed comparison between the latter and the predictions of lukewarm inflation models for different forms of the scalar potential and dissipation coefficient. 

\vspace{0.2cm}

{\bf Acknowledgments:} We thank Rudnei Ramos for helpful discussions. This work was supported by national funds by FCT, through the research projects with DOI identifiers 10.54499/UID/04564/2025, 10.54499/CERN/FIS-PAR/0027/2021, and by the project 10.54499/2024.00252.CERN  funded by measure RE-C06-i06.m02 – ``Reinforcement of funding for International Partnerships in Science, Technology and Innovation'' of the Recovery and Resilience Plan - RRP, within the framework of the financing contract signed between the Recover Portugal Mission Structure (EMRP) and the Foundation for Science and Technology I.P. (FCT), as an intermediate beneficiary. P. B. F. was supported by the FCT - Fundação para a Ciência e Tecnologia, I.P. fellowship SFRH/BD/151475/2021 with DOI identifier 10.54499/SFRH/BD/151475/2021. ATM was supported by the Slovenian Quantum Science Hub co-funded by the Marie Skłodowska-Curie Actions programme (GA-101177446) and the Slovenian Research and Innovation Agency (ARIS), contract number 5110-18/2025-5.

\section*{Supplemental Material}

To better understand the behavior of the PBH mass function at small and large masses, we may firstly note that the integral in Eq.~(\ref{PBH_spectrum}) is dominated by values of the horizon mass $M_H$ for which $\mu^{1/\gamma}< \delta_c$, so that we may restrict the integration to the range $M_H>M\delta_c^{\gamma}/k\equiv M_{H*}$, where the factor $\mu^{1/\gamma}$ in the Gaussian exponential may be discarded. Secondly, we may neglect the variance of density fluctuations outside the interval $M_{H\, min}< M_H< M_{H\, max}$ limited, up to $\mathcal{O}(1)$ numerical factors, by the horizon mass when the scales $k_{max}$ and $k_{min}$ reenter the horizon after lukewarm inflation. Within this horizon mass range we have approximately $\sigma^2\simeq {8\over 81}A_s(M_H/M_{H\, min})^{1-n_s\over 2}$. Since typically $M_{H\, max}\gg M_{H\, min}$, we may also take the limit $M_{H\, max}\rightarrow \infty$. We then have, up to constant factors:
\begin{equation*}
f_{PBH}(M)\sim M^{1+{1\over\gamma}}\int_{\bar{x}}^\infty x^{-{5\over 2}-{1\over \gamma}+{1\over4}(n_s-1)}e^{-ax^{n_s-1\over 2}}~,   
\end{equation*}
where $x=M_H/M_{H\, min}$, $\bar{x}=\textrm{max}(1, M_{H*}/M_{H\, min})$ and $a$ is defined in the main text. On the one hand, for small PBH masses, $\bar{x}=1$, and hence $f_{PBH}\sim M^{1+{1\over\gamma}}$. On the other hand, for large masses the lower integration limit depends on $M$. We may then use ${x^{n_s-1\over 2}\simeq 1+{1\over 2}(n_s-1)\ln x}$ for a nearly scale-invariant spectrum to perform the integral, obtaining $\bar{x}^{-{3\over 2}-{1\over \gamma}-{1\over 4}(2a-1)(n_s-1)}$. For $a\gg {1\over 2}$, combining with the pre-factor $M^{1+{1\over \gamma}}$ we then find $f_{PBH}\sim M^{-{1\over 2}-{a\over 2}(n_s-1)}$ as quoted in the main text. Note that this result is only approximate, but we find that it reproduces the numerical evaluation of $f_{PBH}$ shown in Fig.~\ref{pbhs} up to percent level changes in $n_s$.

 The transition between the two behaviors thus occurs for $M_{H_*}= M_{H\, min}$, corresponding to ${M=M_* \sim4\pi (90/\pi^2 g_*)^{1/2} K\delta_c^\gamma M_P^3/T_R^2}$, up to the $\mathcal{O}(1)$ numerical uncertainty associated with the mass scale $M_{H\, min}$. This can be inferred from the numerical results for $f_{PBH}(M)$, yielding the expression for $M_*$ quoted in the main text.

\vfill
%%%%%%%%%%%%%%%%%%%%%%%%%%%%%%%%%%%%%%%%%%%%%%%%%%%%%%%%%%%%%%%%%%%%%%%
%%%%%%%%%%%%%%%%%%%%%%%%%%%%%%%%%%%%%%%%%%%%%%%%%%%%%%%%%%%%%%%%%%%%%%%
%%%%%%%%%%%%%%%%%%%%%%%%%%%%%%%%%%%%%%%%%%%%%%%%%%%%%%%%%%%%%%%%%%%%%%%
%%%%%%%%%%%%%%%%%%%%%%%%%%%%%%%%%%%%%%%%%%%%%%%%%%%%%%%%%%%%%%%%%%%%%%%

\vspace{0.2cm}

%\bibliography{Superradiant_DM.bib}

\begin{thebibliography}{99}
\bibitem{footnote1}
 We note that in other scenarios the dissipation coefficient may exhibit a steeper temperature dependence, with the constant $C_T$ generically replaced by a ratio $(T/M)^n$ between the radiation temperature and a large mass scale $M$ characteristic of each setup, with the power $n$ also depending on the nature of the radiation fields.
\bibitem{footnote2}
 The discrete interchange symmetry of the model translates into a $\mathbb{Z}_2$ reflection symmetry for the inflaton field, but this may or may not be imposed as discussed.
\bibitem{footnote3}
Here we neglect the mild dependence on the number of relativistic degrees of freedom when the overdense region reenters the horizon, since $k^{-1}=(aH)^{-1}\sim g_*(T)^{-1/6}$.
\end{thebibliography}

\end{document}